\documentclass[11pt,preprint]{aastex}

\begin{document}

\title{Prompt Mergers of Neutron Stars with Black Holes}

\author{M. Coleman Miller}
\affil{Department of Astronomy, University of Maryland\\
       College Park, MD  20742-2421\\
       miller@astro.umd.edu}

\begin{abstract}

Mergers of neutron stars with black holes have been suggested as
candidates for short gamma-ray bursts.  They have also been
studied for their potential as gravitational wave sources
observable with ground-based detectors.  For these purposes, it is
important to know under what circumstances such a merger could
leave an accretion disk or result in a period of stable mass
transfer. We show that, consistent with recent numerical
simulations,  it is expected that mergers between neutron stars
and black holes will be prompt, with no accretion disk and no
stable mass transfer, if the black hole has a mass greater than
that of the neutron star and is spinning slowly.  The reason is
that for comparable masses, angular momentum loss to gravitational
radiation starts a plunge orbit well outside the innermost stable
circular orbit, causing direct merging rather than extended mass
transfer. Even when the black hole is spinning rapidly and exactly
prograde with respect to the orbit, we show that it is possible
within current understanding that no accretion disk will form under
any circumstances, but resolution of this will require
full general relativistic numerical simulations with no approximations.

\end{abstract}


\section{Introduction}

Compact object mergers are expected to be a prominent source for
ground-based gravitational wave detectors (e.g., Nutzman et al.
2004), and have also been invoked as the central engines of short
gamma-ray bursts (e.g., Narayan, Paczynski, \& Piran 1992;
M\'esz\'aros, Rees, \& Wijers 1999; Piran 1999; Fryer, Woosley, \&
Hartmann 1999).  There has therefore been significant effort devoted
to numerical modeling of such mergers. Most of this effort has focused
on mergers between two black holes or two neutron stars, but as black
hole -- neutron star coalescence could be a prominent source of
high-frequency gravitational waves there have also been some studies
of this type of event.

The earliest simulations used a Newtonian potential for simplicity
(Klu\'zniak \& Lee 1998; Uryu \& Eriguchi 1999; Lee \& Klu\'zniak
1999a,b; Janka et al. 1999; Lee 2000, 2001;  Rosswog, Speith, \& Wynn
2004).  These simulations typically showed a period of stable mass
transfer from the neutron star to the black hole, in which the widening
of the orbit caused by the transfer is combined with loss of angular
momentum to gravitational radiation (other papers discussing this idea
include Clark \& Eardley 1977; Jaranowski \& Krolak 1992; Portegies
Zwart 1998; and Davies, Levan, \& King 2005). As a result, the merger is
extended greatly in duration.  Some of these simulations even showed a
``bouncing" effect, by which the onset of mass transfer produces a
significant eccentricity in the orbit and mass is donated in bursts
rather than continuously.  If this were a realistic description of such
mergers, it would have a major effect on the gravitational radiation
waveforms and electromagnetic signatures of these events.

However, as we discuss here, Newtonian simulations are {\it not}
realistic, and in fact give a qualitatively incorrect picture of the
merger. Indeed, some recent numerical simulations of BH-star mergers
that include general relativistic pseudopotentials (e.g., Klu\'zniak
\& Lee 2002; Rosswog 2005) or motion of a neutron star in the
background spacetime of a massive black hole (Rasio et al. 2005) show
that in many circumstances there is simply a direct merger, rather
than stable mass transfer or the formation of an accretion disk.
This was anticipated by Kochanek (1992; see the end of his section
7), and has dramatically different implications for short gamma-ray
bursts and gravitational waves.

Here we show that prompt mergers are a natural result of general
relativistic effects that have no Newtonian analog.  As emphasized
first in this context by Rasio \& Shapiro (1999; see their pages
18-19), of special importance is the existence of a minimum in the
specific angular momentum of circular orbits, at the innermost stable
circular orbit (ISCO). Only if tidal disruption occurs outside the ISCO
can there be an accretion disk or stable mass transfer; otherwise, the
matter simply plunges directly into the black hole.  The ratio of tidal
radius to ISCO radius decreases with increasing black hole mass,  hence
only low-mass black holes can disrupt neutron stars outside the ISCO.
In this limit, however, other effects enter to enhance the likelihood
of a plunge.  First, if the mass of the neutron star is non-negligible,
then the dynamical instability of the orbit occurs at a greater
separation than the radius of the ISCO computed using just the mass of
the black hole.  Second, the flatness of the specific angular momentum
profile near the ISCO means that even a small loss of angular momentum
will lead to a rapid plunge.  For a comparable-mass BH-NS binary,
gravitational radiation is a significant sink of angular momentum.  As
a result, the effective radius of dynamical instability is appreciably
outside the ISCO, and a one-time plunge becomes possible even for
low-mass black holes, if the spin parameter of the black hole is modest.

In \S~2 we expand quantitatively on this argument for non-spinning
black holes, using recent semi-analytic results.  In \S~3 we discuss
the role of the spin.  We show that although encounters
with rapidly spinning black holes can result in accretion disks
and possibly stable mass transfer, the parameter space for this
is surprisingly small because for a comparable-mass binary in
which one component (the neutron star) is spinning slowly, the
effective spin of the system is reduced substantially and hence
the radius of the ISCO is increased greatly.  We discuss the
implications of our findings in \S~4.

\section{Non-Rotating Black Holes}

To determine whether tidal disruption occurs inside or
outside the ISCO, we need to compute (1)~the separation at
which tidal stripping begins to occur, (2)~the separation for
which dynamical inspiral begins (i.e., the effective separation
at the ISCO), and (3)~the effects of gravitational radiation on
the plunge.  The first two points will require numerical
simulations for full understanding, but as a guide we will use
recent semi-analytic results.  For the last point we will
apply the lowest-order quadrupolar radiation expressions
derived by Peters \& Mathews (1963) and Peters (1964), to
get a rough idea of the magnitude of the effect.

In this section and the rest of the paper, we assume that a neutron
star of gravitational mass $m_{\rm NS}$ orbits a black hole of
gravitational mass $m_{\rm BH}$, with an orbital separation $r$.
Note that to be strictly correct we would need to state precisely
in which coordinate system we measure $r$, but for the estimates in
this paper we will assume that $r$ is roughly a Boyer-Lindquist
radius even though for a comparable-mass binary the spacetime will
not be Kerr.  With these masses we define a total mass $M\equiv
m_{\rm NS}+m_{\rm BH}$, a reduced mass $\mu\equiv m_{\rm NS}m_{\rm
BH}/M$, and a symmetric mass ratio $\nu\equiv\mu/M$.  Note that the
maximum value of $\nu$ is 1/4; this makes it a good expansion
parameter, a feature exploited in various post-Newtonian (e.g.,
Blanchet 2002) and equivalent one-body (Buonanno \& Damour 1999;
Damour, Jaranowski, \& Sch\"afer 2000; Damour 2001) treatments of
strong gravity.  We also assume that the  dimensionless spin
parameter of the black hole is ${\hat a} \equiv J/m_{\rm BH}^2$ (in
geometrized units in which $G=c=1$, which  we use henceforth;
${\hat a}=0$ for this section only) and that the equilibrium radius
of the neutron star far from tidal fields is $R_0$.  The lack of
sufficient viscosity to enforce corotation (Kochanek 1992; Bildsten
\& Cutler 1992) suggests that the neutron star will be rotating
slowly.

We now treat in order the tidal stripping, the separation at
the ISCO, and the effects of gravitational radiation.

\subsection{Radius of tidal stripping}

Several authors have generalized the Newtonian fluid analysis of
tidally locked ellipsoids to general relativistic orbits (e.g.,
Fishbone 1973; Mashoon 1975; Wiggins \& Lai 2000).  Here we follow the
treatment of Wiggins \& Lai (2000), who analyze the physically more
realistic irrotational case as well as the corotating orbits examined
in previous treatments (note that, as shown by Bildsten \& Cutler 1992,
the weaker tidal fields from more massive objects imply that it is
even more difficult to synchronize a BH-NS system than a NS-NS system).

Wiggins \& Lai (2000), like other authors, simplify by assuming
that $m_{\rm BH}\gg m_{\rm NS}$, but as is standard we will proceed
with caution by applying their formulae even to comparable-mass
binaries.  We note that even in the limit of equal mass objects,
their results are reasonable: for example, they predict that in
this case mass transfer would begin when the separation is slightly more
than twice the unperturbed radius of the neutron star, which is
expected because the NS will be distorted in the direction
of the black hole.  They also note that their treatment of the neutron star's
self-gravity as Newtonian (a standard assumption in such work)
actually underestimates its binding, because at $m_{\rm NS}/R_0\sim 0.2$,
general relativistic gravity is significantly stronger than Newtonian
gravity.  The effective tidal radii are therefore actually smaller
than we quote here, hence our conclusions are conservative (that is,
neutron stars are even more likely to merge directly without
forming an accretion disk).

Wiggins \& Lai (2000) define two dimensionless parameters,
\begin{equation}
{\hat R}_0\equiv {R_0\over m_{\rm NS}}
\left(m_{\rm NS}/m_{\rm BH}\right)^{2/3},
\qquad {\hat r}\equiv (r/m_{\rm BH})/{\hat R}_0\; ,
\end{equation}
to characterize the effect of tides.  When ${\hat r}$ is less than
some critical value ${\hat r}_{\rm tide}$, tidal stripping begins.
For ${\hat a}=0$ they find 
${\hat r}_{\rm tide}\approx 2-2.4$, depending on the polytropic
index $n$ and only very weakly dependent on whether the neutron
star is corotating or irrotational (Wiggins \& Lai 2000 show that
for corotating binaries ${\hat r}_{\rm tide}$ is larger, but only
by $\sim$3\%, than it is for irrotational binaries).  Given a value of 
${\hat R}_0$, one can then compute the actual separation
$r$ at which the stripping occurs.  For a particular critical
$r/M$ (e.g., $r/M=6$, the radius of the ISCO in the Schwarzschild
spacetime), one can therefore define ${\hat R}_{0,{\rm crit}}$
such that for ${\hat R}<{\hat R}_{0,{\rm crit}}$ disruption occurs
inside the critical radius.  For $r_{\rm crit}/M=6$ and ${\hat a}=0$, 
they find ${\hat R}_{0,{\rm crit}}=2.54$ for $n=1$ and 
${\hat R}_{0,{\rm crit}}=2.76$ for $n=3/2$.  By itself this would
suggest that for a neutron star of mass $m_{\rm NS}=1.4\,M_\odot$
and radius $R_0=5m_{\rm NS}$ (approximately 10~km), a black hole of mass
greater than $\approx 4.4\,M_\odot$ would swallow the star whole.
As we show in the next two sections, however, other effects
lower this threshold significantly.

\subsection{Radius of the ISCO}

Interest in the gravitational wave signatures of merging binary black
holes has led to the development of schemes to approximate the
spacetime when the two objects in the binary have comparable mass
(e.g., Buonanno \& Damour 1999; Damour et al. 2000; Damour 2001;
Blanchet 2002).  These usually involve expansions to some
post-Newtonian (PN) order, where an nPN expansion means including
terms out to $(v^2/c^2)^n$.  As tidal effects enter only at the 5PN
order (e.g., Wiggins \& Lai 2000 and many other references), they can
largely be ignored.

Several authors have used these schemes to estimate the separation
of a binary at the ISCO for arbitrary mass ratios.  For zero spin
and two equal masses ($\nu=1/4$), they find $r_{\rm ISCO}\approx 5M$
(see Damour et al. 2000, equations 4.36c, 4.40c, and 4.41 for the
formulae needed to compute angular momenta for circular orbits with
two arbitrary masses; note that their parameter $\omega_{\rm static}=0$, as is
determined in Damour, Jaranowski, \& Sch\"afer 2002).
Recalling that $M=m_{\rm NS}+m_{\rm BH}$, this means that the
separation is significantly greater than the $6m_{\rm BH}$ separation
one would obtain by just considering one black hole.  The symmetric
mass ratio is relatively flat for comparable masses (e.g., for 
$m_{\rm BH}=2m_{\rm NS}$ we have $\nu=2/9$), so 
$r_{\rm ISCO}\approx 5M$ holds for all mass ratios in this range.
Therefore, let us reconsider the critical black hole mass such
that a neutron star is disrupted outside the ISCO.  For
$m_{\rm BH}/m_{\rm NS}=2$ and $R_0=5m_{\rm NS}$, 
${\hat R}_0=3.15$.  At this ${\hat R}_0$, Wiggins \& Lai (2000) find
${\hat r}_{\rm tide}\approx 3$ and hence $r/m_{\rm BH}=7.25$.
Therefore, $r/M=4.83$, and this orbit is actually inside the ISCO.
Consider now a minimal mass black hole of $m_{\rm BH}=2.2\,M_\odot$
in a binary with a neutron star of $m_{\rm NS}=1.4\,M_\odot$ and
$R_0=5m_{\rm NS}$.  Then ${\hat R}_0=3.70$, ${\hat r}_{\rm tide}=
2.27$ according to Wiggins \& Lai (2000) for $n=1$, and thus
$r/m_{\rm BH}=8.40$.  This implies $r/M=5.13$.  This is technically
just outside the ISCO, but as we will see in the next section,
angular momentum loss to gravitational radiation is significant
and the plunge actually starts well outside this radius.  Higher
mass neutron stars are more compact (with $R_0<5m_{\rm NS}$), and will
also not be tidally disrupted.  Only low mass stars with
$R_0>5m_{\rm NS}$ might leave accretion disks, but as we now show, even
this is not certain.

\subsection{Effect of gravitational radiation losses}

The ISCO as a sharp dividing line is only strictly useful in the test
particle limit, where no other effects could cause inspiral.  Any
mechanism that leads to inspiral of nearly circular orbits, whether
magnetic angular momentum transport in accretion disks or losses to
gravitational radiation, will blur this line (see, e.g., Buonanno \&
Damour 2000 for a detailed discussion of this phenomenon).  Near to
but outside of the ISCO, the specific angular momentum of circular
orbits is very flat with radius, meaning that even a small amount of
angular momentum loss can have drastic consequences (see, e.g.,
Lombardi, Rasio, \& Shapiro 1997).  In particular,
even after the neutron star has reached the tidal radius, the matter
that is stripped will take some finite time to separate significantly
from the star, and during this time all the matter is still emitting
gravitational radiation. We suspect that the omission of these effects
by Prakash, Ratkovic, \& Lattimer (2004), who did a semianalytic
treatment of BH-NS mergers using the Blanchet (2002) post-Newtonian
equations, is the reason that they reached qualitatively different
conclusions from ours.

Consider, for example, loss to gravitational radiation.  From Peters
\& Mathews (1963) and Peters (1964), the lowest-order (quadrupolar)
loss rate of angular momentum for a nearly circular orbit is
\begin{equation}
dL/dt=-{32\over 5}(r/M)^{-7/2}\nu\mu\; .
\end{equation}
Let the angular momentum at the ISCO be
$L_{\rm ISCO}$.  If enough angular momentum is lost
in a time short compared to the orbital period so that 
$L<L_{\rm ISCO}$, then the black hole and neutron
star will plunge together.  For the hypothetical 
$1.4\,M_\odot - 2.2\,M_\odot$ binary discussed above, an
initial circular orbit of radius $5.13\,M$ has an
angular momentum only $0.00078\mu M$ greater than that of the ISCO.  
From the formula above
for angular momentum loss, this is radiated within a time
$0.16M$, compared to an orbital period of $\approx 75\,M$ at
that radius.  Clearly, the binary undergoes a plunge well
before the neutron star is tidally stripped.

These effects are summarized in Figure~1.  Again assuming no
spin of the black hole or neutron star, we show the critical
radius to mass ratio for a neutron star, as a function of the
ratio of the black hole mass to neutron star mass.  The solid
lines are for different criteria for the plunge, including the
effects of gravitational radiation.  For comparison we show
$R_0/m_{\rm NS}$ versus mass ratio for neutron stars with hard
and soft equations of state, assuming a black hole with a mass
just above the neutron star maximum in each case (to maximize
the likelihood of tidal disruption outside the ISCO).  We see
that the neutron star is likely to be disrupted inside the
radius of dynamical instability.  Therefore, if the black hole
is only spinning slowly, it will eat a neutron star whole
rather than forming an accretion disk or producing stable mass
transfer.

\section{Effects of Black Hole Spin}

Determination of the effective separation at the ISCO is much
more challenging when the spin is nonzero and orbits are
prograde.  This is because the separation decreases with
increasing spin, hence post-Newtonian approximations get
worse and full numerical relativity may be necessary.
Nonetheless, we can reach some tentative conclusions 
following the work of Damour (2001) in constructing
effective one-body spacetimes for comparable masses
including spins.  The critical point here is that although
in principle the radius of the ISCO could be as low as $M$
for prograde orbits in a maximally rotating Kerr spacetime, this
radius plunges dramatically at high spin, hence if the effective
spin is slightly less than maximal then the ISCO radius is much larger
and it becomes more difficult to disrupt neutron stars before
merger.  This also applies for orbits that are not exactly prograde,
for which the effective radius of the ISCO is increased dramatically
for rapidly spinning black holes (see Hughes \& Blandford 2003).

Damour finds that for two bodies of
masses $m_1$ and $m_2$ and spin angular momenta
${\bf S}_1$ and ${\bf S}_2$, the effective angular momentum
of the one-body spacetime is
\begin{equation}
{\bf S}^{\rm eff}=\left(1+{3\over 4}{m_2\over m_1}\right){\bf S}_1
+\left(1+{3\over 4}{m_1\over m_2}\right){\bf S}_2\; .
\end{equation}
The effective dimensionless spin parameter is then
${\hat a}_{\rm eff}={\bf S}^{\rm eff}/M^2$.  
Because the neutron star is not expected to be rotating rapidly
(that is, ${\bf S}_{\rm NS}\approx 0$),
the effective spin of the spacetime is diminished, which also means
that the merger properties are not as sensitive to the precise black
hole spin as they would be if the neutron star mass were negligible.
For example, even if the black hole is rotating maximally
(${\bf S}_{\rm BH}=m_{\rm BH}^2$), the maximum ${\hat a}_{\rm eff}$ for
an equal-mass binary is only ${\hat a}_{\rm max}=0.44$ and
for $m_{\rm BH}=2m_{\rm NS}$ is only ${\hat a}_{\rm max}=0.61$.

Unfortunately, the approximations used by Damour (2001) get
progressively worse with larger ${\hat a}_{\rm eff}$, so it is difficult to
draw definite conclusions.  The abrupt shift in the ISCO near ${\hat
a}_{\rm eff}=1$ is likely to mean that with a comparable-mass binary the ISCO
is well outside its theoretical minimum radius of $M$; for example,
in a Kerr spacetime
$r_{\rm ISCO}=4.46\,M$ when ${\hat a}_{\rm eff}=0.44$, and $r_{\rm
ISCO}=3.79\,M$ when ${\hat a}_{\rm eff}=0.61$ (for the relevant expressions,
see, e.g., Shapiro \& Teukolsky 1983, chapter 12).  The exact numbers
will change for objects of comparable mass, but we expect the
qualitative effect to be similar: the effective spin will not be near
extremal for comparable-mass objects, so dynamical instability may
still set in prior to tidal stripping.

Counterintuitively, therefore, somewhat {\it higher} mass ratios
could be more likely to result in accretion disks and stable mass
transfer for a rapidly spinning black hole, because the effective
spin could be closer to maximal.  It might, however, still be
difficult to generate stable mass transfer.  For example,
consider a extreme Kerr black hole  of mass $m_{\rm
BH}=7\,M_\odot$ orbited by a neutron star of mass $m_{\rm
NS}=1.4\,M_\odot$ and radius $R_0=5m_{\rm NS}$.  The higher mass
ratio means that the full spacetime is likely to be reasonably
close to Kerr, making extrapolations more reliable.  We have
${\hat R}_0=1.71$ and ${\hat r}_{\rm tide}=2.3$. Then $r/m_{\rm
BH}=3.93$, so $r/M=3.28$.  Using the Damour (2001)
approximation, ${\hat a}_{\rm eff}=0.80$, so in the Kerr spacetime $r_{\rm
ISCO}/M=2.91$.  Taking these numbers at face value, even in this
case gravitational radiation losses would cause the neutron star
to plunge before it was stripped.

However, this is close enough to the tidal radius that details
are important: for example, if in reality the effective spin parameter is
${\hat a}_{\rm eff}=0.90$, then $r_{\rm ISCO}/M=2.32$ and stable
mass transfer might ensue.  We are therefore
in a domain where existing approximations are inadequate to
determine the fate of the merging neutron star.  Full
no-approximation general relativistic calculations will be
necessary.  Nonetheless, it appears possible that {\it no}
plausible combination of neutron star and black hole masses,
spins, and orbital inclinations will result in an accretion disk 
or stable mass transfer.

\section{Conclusions}

Mergers of neutron stars with black holes are of great interest for
models of short gamma-ray bursts and as sources of high-frequency
gravitational radiation.  The likely signatures of such events depend
strongly on whether the star is swallowed whole in a direct merger or
whether it is tidally stripped far enough away from the black hole
that an accretion disk forms or stable mass transfer occurs.  We have
shown that for slowly rotating black holes, current calculations of
tidal stripping and of the spacetimes of binary compact objects
suggest that direct merger will almost always take place. This is in
agreement with recent general relativistic numerical simulations,  and
could actually enhance the likelihood of a baryon-free environment for
a short gamma-ray burst (e.g., M\'esz\'aros \& Rees 1992).  A prograde
encounter of a neutron star with a rapidly rotating black hole is more
likely to result in a disk, but the current understanding of such
close mergers is uncertain enough that it could be that {\it any}
NS-BH merger will be direct.  Future fully general relativistic
numerical simulations will be required to resolve this issue.

\section*{Acknowledgments}
We thank Alessandra Buonanno, Melvyn Davies, and especially the
referee Fred Rasio for helpful comments.
We also thank the Theoretical Institute for Advanced Research in
Astrophysics (Hsinchu, Taiwan) for hospitality during part of this
work.  This paper was supported in part by NASA grant NAG 5-13229.

\newpage

\figcaption[]
{Critical radius to mass ratio for a neutron star, as a function of
black hole to neutron star mass ratio, for no rotation.  Below the
critical value, the neutron star is compact enough to plunge and be
swallowed whole rather than be disrupted.  The top solid line is
constructed by assuming that the neutron star will plunge when, in
one full orbit, it can reduce its angular momentum below the ISCO
value via emission of gravitational radiation.  The next two solid
lines down reduce the allowed time to 30\% and 10\% of an orbit,
respectively, and the bottom line ignores gravitational radiation
losses entirely.  For comparison, we show $R_{\rm NS}/m_{\rm NS}$ versus
$m_{\rm BH}/m_{\rm NS}$ for n-body equations of state (Akmal et al. 1998)
that are hard (A18+UIX+dvb, in the Akmal et al. 1998 notation;
dashed line)  and soft (A18; dotted line).  Stars constructed with
relativistic mean field theory could be somewhat larger (see
Lattimer \& Prakash 2001 for a recent review).  In both cases we
assume that $m_{\rm BH}$ is just above the maximum mass of a neutron
star, because this maximizes the likelihood of tidal disruption.  We
also consider  $m_{\rm NS}$ as small as $1.25\,M_\odot$, which is the
lowest gravitational mass yet measured for a neutron star (for
PSR~J0737--3039B; see Lyne et al. 2004).  This figure shows that
direct merger, not stable mass transfer, is the likely outcome of
the coalescence of a neutron star with a nonrotating black hole.
}

\newpage

\centerline{\plotone{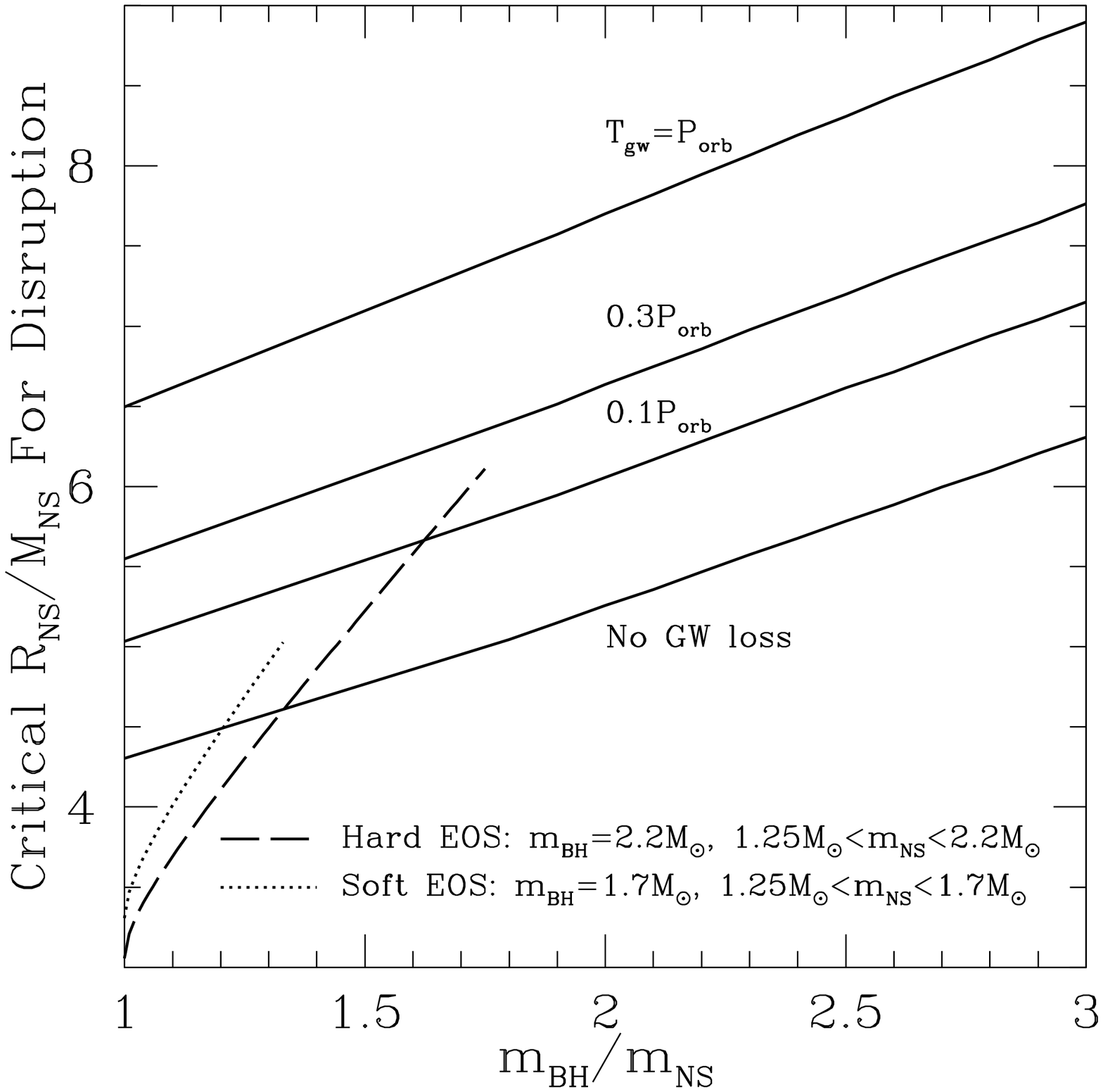}}

\end{document}